\def\Granadadep{Departamento de F\'\i sica Te\'orica y del Cosmos, Facultad
de Ciencias, Universidad de Granada, Campus de Fuentenueva, Granada 18002,
Spain}

\def\Granadainst{Instituto de F\'\i sica Te\'orica y Computacional Carlos I, 
Facultad
de Ciencias, Universidad de Granada, Campus de Fuentenueva, Granada 18002, 
Spain}

\def\Valencia{IFIC, Centro Mixto Universidad de Valencia-CSIC, Burjasot
              46100-Valencia, Spain.}
\def\Comision{Work partially supported by the DGICYT.}

\def\nn{\nonumber}
\def\ni{\noindent}

\def\be{\begin{equation}}
\def\ee{\end{equation}}
\def\bea{\begin{eqnarray}}
\def\eea{\end{eqnarray}}
\def\ba{\begin{array}}
\def\ea{\end{array}}

\def\z{\zeta}

\def\ep{\epsilon}

\def\z{\zeta}

\def\ni{\noindent}

\newcommand{\parcial}[1]{ \frac{\partial}{\partial #1} }

\newcommand{\XL}[1]{ {\tilde{X}}^{L}_{#1} }

\newcommand{\XR}[1]{ {\tilde{X}}^{R}_{#1} }

\documentstyle[11pt]{article}

\begin{document}
 


\hbox{ }

\vskip 3 cm

\begin{center} 
{\Large {\bf ALGEBRAIC VERSUS TOPOLOGIC ANOMALIES$^1$ }}
\end{center}

\bigskip
\bigskip

\centerline{ V. Aldaya$^{2,3}$, M. Calixto$^2$ 
              and J. Guerrero$^{2,4}$  }       

\bigskip

\footnotetext[1]{\Comision}
\footnotetext[2]{\Granadainst} \footnotetext[3]{\Valencia}
\footnotetext[4]{\Granadadep}  
\footnotetext{This paper is in final form and no version of it will be
      submitted for publication elsewhere} 

\bigskip

\begin{center}
{\bf Abstract}
\end{center}

\small

\begin{list}{}{\setlength{\leftmargin}{3pc}\setlength{\rightmargin}{3pc}}
\item Within the frame of a Group Approach to Quantization anomalies 
arise in a quite natural way. We present in this talk an analysis 
of the basic obstructions that can be found when we try to translate
symmetries of the Newton equations to the Quantum Theory. They fall
into two classes: {\it algebraic} and {\it topologic} according to the 
local or global character of the obstruction. We present here one 
explicit example of each.
 
\end{list}

\normalsize

\vskip 1cm

\section{The concept of anomaly}
Roughly speaking  {\it anomalies} are {\it obstructions} to the {\it 
quantum realization} of  {\it classical symmetries} and they fall into 
two classes according to the local or global character of the obstruction.
On the one hand, there are generators in the symmetry group that do not
preserve any distinction between the basic $\hat{q}$ and $\hat{p}$ 
quantum operators, and this pathology shows up at the Lie algebra level.
We shall call this obstruction {\it algebraic anomaly}. On the other hand,
we can find symmetry generators, the local action of which (in the sense of local chart)
is well-behaved, whereas their finite action, the exponential, does not 
preserve the Hilbert space of the theory. We shall call this type of 
obstruction {\it topologic anomaly} and those generators, {\it bad operators}.

Algebraic anomalies firstly appeared in Quantum Field Theory when trying
to quantize a current algebra made of fermionic matter \cite{Jackiw}. Let $G$ be a group
of internal symmetries of a classical relativistic theory, generated by
$T_a$ such that

\begin{equation}
\left[T_a,T_b\right]=C^c_{ab}T_c 
\end{equation}

\ni The classical Lagrangian theory provides Noether currents $j_a\equiv *i_{T_a}\Theta$
($j^{\mu}_a=\frac{\partial{\cal L}}{\partial \psi_{\mu}}T_a)$
so that the equation $\partial_\mu j^{\mu}_a=0$ implies the conservation
of the charges $Q_a\equiv \int_{\Sigma}d\sigma_{\mu}j^{\mu}_a$, which
close the Lie algebra ${\cal G}$ of $G$ under the Poisson bracket:

\begin{equation}
\{Q_a,Q_b\}=C^c_{ab}Q_c \,\, .
\end{equation}

Occasionally, we can formally apply the Poisson bracket to $j^{\mu}_a$ itself, even
though it is not a Noether invariant, and find a closed current algebra:

\begin{equation}
\{j^0_a(\vec{x}),j^i_b(\vec{y})\}=C^c_{ab}\delta(\vec{x}-\vec{y})j^i_a,
\,\,\Sigma\equiv\{x^0=cte\} \;\; ,\label{current}
\end{equation}

\ni thus interpreting this algebra as a classical symmetry.

The problem then arises of up to what extent the algebra (\ref{current})
can be quantized {\it in terms of the basic quantum} (fermionic) {\it operators}
of the theory. One usually finds, in fact, {\it extra (Schwinger) terms}
on the r.h.s. of the quantum version of (\ref{current}), which are referred to as {\it
anomalous terms}.

A more precise example is that of {\it conformal symmetry} in {\it string
theory} (see, e.g. \cite{Goddard}). Here the classical conformal symmetry is unambiguously defined. The
classical generators $L_m$ are written in terms of the classical normal modes
$\alpha_m$ (the {\it basic} variables) of the string:

\begin{equation}
L_m=\sum_n\alpha_n\alpha_{m-n} \;\; .
\end{equation}

\ni $L_m$ are true Noether invariants satisfying the Poisson algebra:

\begin{equation}
\{L_m,L_n\}=(m-n)L_{m+n}\,\, .
\end{equation}

\ni However, the quantization procedure allows $\hat{L}_n$ to be written
in terms of the {\it basic operators} $\hat{\alpha}_m$ {\it only} for a 
particular value of the deformation parameter of the conformal algebra:

\begin{equation}
[\hat{L}_m,\hat{L}_n]=(m-n)\hat{L}_{(m+n)}+\frac{(d=26)}{12}n^3\delta_{m,n}\hat{1} \;\; .
\end{equation}

\ni For different values of $d$, $\hat{L}$ cannot be written in terms of
the $\hat{\alpha}$'s and, therefore, $\hat{L}$ {\it must also be considered as
basic operators} of the theory.

{\it Topologic anomalies} have been characterized by the failure of the
Ehrenfest theorem in quantum systems formulated on a configuration space 
with non-trivial topology \cite{Manton,Frachal}:

\begin{equation}
\frac{d\,}{dt}<\hat{A}>=\frac{i}{\hbar}<[\hat{H},\hat{A}]>+anomaly\;\;. 
\end{equation}

\ni This is a {\it global} problem {\it without} classical counterpart
provided that Classical Mechanics is formulated by {\it initial value problems},
and it is often related to the lack of hermiticity of quantum operators.

Topologic anomalies also appear along with {\it non-equivalent quantizations}
of the starting classical system, the phase space of which has {\it non-trivial
homotopy group} $\pi^1$ \cite{Landsman} (see also \cite{Souriau}.

\section{Anomalies in a Group Approach to Quantization}
Anomalies are more accurately formulated in a framework where symmetry
and quantization are strongly related issues. We shall discuss them in an
already formulated Group Approach to Quantization \cite{23,Ramirez}.

On a Lie group we have two types of {\it mutually commuting} operators:
Left- and Right-invariant vector fields
\begin{itemize}
\item{} Right-invariant fields are {\it pre-quantum operators} providing a
        {\it reducible} representation
\item{} Left-invariant fields are used to {\it reduce} the representation 
        in a compatible way
\end{itemize}

\ni In addition, the quantization group $\tilde{G}$ is supposed to wear a 
principal bundle structure with fibre a group $T$ containing $U(1)$:

\begin{equation}
\tilde{G},T\rightarrow \tilde{G}/T\,\, ;
\end{equation}

\ni the subgroup $U(1)$ realizes the ordinary phase invariance in Quantum Mechanics.
and establishes the basic canonically conjugate (symplectic)
variables: those producing a $U(1)$-term on the r.h.s. of their commutator.

For the sake of simplicity let us assume that $\tilde{G}$ is a $U(1)$-central
extension with Lie algebra co-cycle

\begin{equation}
\Sigma:{\cal G}\times{\cal G}\rightarrow R\;\; .
\end{equation}

\ni The {\it kernel of} $\Sigma$, ${\cal G}_{\Sigma}$, is made of
{\it non-symplectic generators}.

We start from $U(1)$-functions on $\tilde{G}$ ($\Xi\psi=i\psi,\,\Xi\in u(1)$)
on which we impose the {\it polarization conditions}

\begin{equation}
\tilde{X}^L\psi=0\;\;\forall \tilde{X}^L\in {\cal P}\;\; ,
\end{equation}

\ni where ${\cal P}$, the polarization, is a maximal left subalgebra
{\it containing} a proper subalgebra ${\cal A}\subset{\cal G}_{\Sigma}$ and 
{\it excluding} $\Xi$.
It must be stressed that a {\it full} polarization, i.e. containing the
whole ${\cal G}_{\Sigma}$ {\it might not} exist.

On polarized $U(1)$-functions the right-invariant vector fields act
properly and {\it irreducibly} defining the operators of the {\it true
quantization}. Those operators whose left counterpart are in ${\cal A}\subset
{\cal G}_{\Sigma}$ can be {\it written} as a {\it function} of the {\it
basic ones}.

If the subgroup $T$ is bigger than $U(1)$, all generators of $T$ different from
$\Xi$ provide {\it constraints}. In that case, {\it not all} right-invariant
vector fields are {\it good} operators since constraints are imposed on the
right. If {\cal H} is the space of $T$-{\it functions}, i.e. functions satisfying:

\begin{equation}
{\tilde{X}^R}_a\psi=D^{(\epsilon)}(T_a)\psi\,\, ,
\end{equation}

\ni where $D$ is a particular representation of $T$, the index $\epsilon$
of which characterizes the quantization, good operators must preserve this
space. The subgroup $G_{\rm good}$ satisfying either of the following conditions
involving the group commutator \cite{Ramirez,Frachal}:

\begin{equation}
Ad(\tilde{G})\left[T,G_{\rm good}\right]\subset G_{\cal P} \;\; {\rm or}
     \;\; [G_{\rm good},T]\subset {\rm ker} D(T)
\end{equation}

\ni is made of good operators.

\subsection*{Algebraic anomalies}
They appear when a {\it full} polarization {\it does not exist} and only
a polarization containing a proper subalgebra ${\cal A}$ of ${\cal G}_{\Sigma}$
can be found \cite{Loll}.

A non-full polarization causes those operators in ${\cal G}_{\Sigma}\backslash{\cal A}$,
the anomalous operators, {\it not to be expressible in terms of the 
basic ones}. In other words, anomalous operators behave as {\it if 
they were basic}, thus giving a {\it deformation term} on the r.h.s. 
of (some of) their commutators. Furthermore, the fact that we do not 
impose on the wave functions as many polarization conditions
as it should in order to reduce the quantum representation (half of 
the generators in ${\cal G}_{\Sigma}\backslash{\cal A}$ are absent
from ${\cal P}$) and ${\cal P}$ is nevertheless maximal, creates some sort of
paradox. The solution to the polarization equations results, however, in
a rather natural situation: the quantum representation is {\it reducible but
not completely reducible}. Now, some kind of mechanism of restriction to
the invariant subspace is in order. We then resort to the concept of
{\it higher-order polarization}.  

The definition of a higher-order 
polarization ${\cal P}^{HO}$ is analogous to that of first-order one except for the fact 
that its elements are allowed to belong to the enveloping algebra 
$U{\cal G}^L$. However, the existence of a (full) higher-order polarization
is only guaranteed for particular values of the deformation parameters,
{\it the quantum values of the anomaly}. The solutions to the equations associated
with ${\cal P}^{HO}$ constitute the invariant subspace in the quantum
representation, which appears for those particular values of the anomaly.

\subsection*{Topologic anomalies}
They appear when the configuration space of the physical system is of the
form $R^n/\check{T}$ where $\check{T}$ is a (probably discrete) ``surgery" group.
The quantization of this system via our group-theoretical method is achieved
by considering as quantization group that of the corresponding unconstrained
system (with configuration space $R^n$) but with fibre a group $T$
such that $\check{T}\equiv T/U(1)$.

As mentioned above, when the structure group is bigger than $U(1)$, not
all the operators in the theory are good, i.e. not all the operators
preserve the subspace of $T$-functions. Those operator destroying
the $T$-function property of wave functions are in general called {\it bad operators}
and in the present case are identified with {\it topologic anomalies}.
Among them some can be found which destroy the $T$-function property
in a rather benign way (they are not so bad): they just change the label
associated with the $T$-property.
To be more precise,  if we denote by ${\cal H}^{(\epsilon)}$ the particular 
subspace of the whole set of $T$-functions, {\cal H}, carrying the 
irreducible representation $D^{(\epsilon)}$ of $T$,  bad operators do not
preserve in general the specific properties associated with the $\epsilon$-property
but some of them can occasionally transform wave functions $\psi^{\epsilon}\in{\cal H}^{(\epsilon)}$
into $\psi^{\epsilon'}\in{\cal H}^{(\epsilon')}$. In this way, these not
so bad operators generate {\it quantization-changing transformations}, because
the index $\epsilon$ parameterizes non-equivalent quantizations.

\section{The simplest, yet relevant example of an algebraic anomaly}
The simplest Lie group wearing an algebraically anomalous structure is the
Schr\"odinger group in $1+1$ dimensions, the symmetry group of the quadratic potential in
one dimension \cite{Niederer}:

\begin{equation}
Ax^2+Bx+C
\end{equation}

\ni The classical (extended) Schr\"odinger Lie algebra is realized immediately as
the Poisson algebra generated by either

\begin{equation}
\{1,p,x,p^2,x^2,p\cdot x\}\;\;\;\; or\;\;\;\; \{1,p,x,p^2+x^2,x^2,x\cdot p\}\; ,
\end{equation}

\ni where we have distinguished between two different choices of Hamiltonian
in the same algebra. Each choice corresponds to a different class of
irreducible representations: in the first case the entire group is represented
on the carrier space for the irreducible representations of the Galilei
subgroup, and in the second, the entire group is represented on the carrier
space for the irreducible representations of the Newton subgroup (the harmonic oscillator
symmetry).

For the sake of concreteness we shall limit ourselves to the second case
and adopt the more appropriate notation $x\equiv\frac{1}{\sqrt{C+C^*}}\;,
\; p\equiv\frac{-im\omega}{\sqrt{2m\omega}}(C-C^*)$, associated with the harmonic oscillator.
We then write for the Schr\"odinger Poisson subalgebra the generators:

\begin{equation}
\{1,C,C^*,CC^*,\frac{1}{2}C^2,\frac{1}{2}{C^*}^2\}\equiv\{1,C,C^*,h,z,z^*\}\;\; ,
\end{equation}

\ni where the quadratic functions generating the $sL(2,R)$-subalgebra
that replaces the time of the Newton group, has been denote by a single 
character to mean that they really correspond to linear elements in the 
abstract Lie algebra. The Schr\"odinger group
can also be seen, in this way, as the classical symmetry of the two-photon
laser system with a Hamiltonian of the form:

\begin{equation}
H=CC^*+\alpha C^2+\alpha^*{C^*}^2+\beta C+\beta^*C^*+\gamma\;\; .
\end{equation}

The classical commutators between the Lie algebra generators are easily
derived by computing the Poisson bracket, giving rise to:

\begin{eqnarray}
\{h,C^*\}=C^*\;&\;&\;\{C^*,C\}=1 \nn \\
\{h,C\}=-C\;&\;&\;\{z^*,z\}=-\frac{1}{2}h \nn \\
\{h,z^*\}=2z^*\;&\;&\;\{C^*,z^*\}=0  \label{Poisson} \\
\{h,z\}=-2z\;&\;&\;\{C,z\}=0 \nn \\
\{C,z^*\}&=&\frac{1}{\sqrt{2}}C^* \nn \\
\{C^*,z\}&=&\frac{1}{\sqrt{2}}C\;\;\; . \nn \\
\end{eqnarray}

\ni The last two commutators reveal a non-diagonalizable action of the
$sL(2,R)$ subalgebra, which generates the kernel of the co-cycle, on the rest. 
This action is precisely responsible for an algebraic anomaly. In fact,
there is no full polarization containing the entire $sL(2,R)$ subalgebra.
At most, a non-full polarization can be found:

\begin{equation}
{\cal P}=<\XL{C},\XL{h},\XL{z}>\;\; . \label{non-full}
\end{equation}

\ni Quantizing with the non-full polarization (\ref{non-full}) results in
a breakdown of the naively expected correspondence between $\hat{z},\hat{z}^{\dag}$
operators and the basic ones:

\begin{eqnarray}
\hat{z}&\neq&\frac{1}{2}\hat{C}^2  \nn \\
\hat{z}^{\dag}&\neq&\frac{1}{2}\hat{C}^{\dag^2}\;\; . 
\end{eqnarray}

Unlike in the classical case, the operators $(\hat{h}),\hat{z},
\hat{z}^{\dag}$ behave independently of $\hat{C},\hat{C}^{\dag}$, although they 
generate an irreducible representation of $SL(2,R)$ with Bargmann index 
$k$ \cite{Bargmann}. The operators $\hat{z},\hat{z}^{\dag}$ seem to have dynamical content as 
if they were canonically-conjugate (basic) operators. In fact, their (quantum)
commutator is no longer the homomorphic image of the corresponding Poisson
bracket (see (\ref{Poisson})) but

\begin{equation}
[\hat{z},\hat{z}^{\dag}]=2(\hat{h}+k\hat{1})\;\; ,
\end{equation}

\ni where an extra central term comes out.

However, only for a particular value (the quantum value of the anomaly)
$k=\frac{1}{4}$, the representation of the Schr\"odinger group {\it 
becomes reducible}, although non-completely reducible and, {\it on the 
invariant subspace}, the operators $\hat{z},\hat{z}^{\dag}$ do really express 
as  $\frac{1}{2}\hat{C}^2,\frac{1}{2}\hat{C}^{\dag^2}$.
The invariant subspace is constituted by the solutions
to a second-order polarization which exists only for 
$k=\frac{1}{4}$ \cite{Loll}:

\begin{equation}
{\cal P}^{HO}=<\XL{C},\XL{h},\XL{z}, \XL{z^*}-\alpha(\XL{C^*})^2>\;\; ,
\end{equation}

\ni where the constant $\alpha$ is forced to acquire the value 
$\alpha=\sqrt{2}k$.

\section{The simplest, yet relevant example of topological anomaly:
the free particle on the circle}
The configuration space $(x,t)$ is the direct product $S^1\times R$,
the fundamental group of which is obviously $\pi^1=Z$. The relevant
symmetry to be used as quantization group is the extended 1+1-Galilei 
group fibered by

\begin{equation}
T=U(1)\times\{e_k,\;k\in Z\}\equiv U(1)\times \check{T}\;\; ,
\end{equation}

\ni $\check{T}$ being the subgroup of finite translations on $x$ by an 
amount of $kL$, for some spatial period $L$. As mentioned above, the inequivalent
quantizations when a generalized ``phase" invariance is involved, are characterized
by the irreducible representations of the structure group $T$:

\begin{eqnarray}
{\cal D}^{(\epsilon)}(\zeta,e_k)&=&\zeta D^{(\epsilon)}(e_k)\;,\;\;\zeta\in U(1) \\
D^{(\epsilon)}(e_k)&=&e^{\frac{i}{\hbar}\epsilon kL}\;\,\;\;\epsilon\in
          [0,\frac{2\pi\hbar}{L})\;\; . \nn 
\end{eqnarray}

Since the non-triviality of the topology comes from the spatial variable
only, let us forget about time and consider the Heisenberg-Weyl
group parameterized by $(x,p,\zeta)$ with a specially suitable co-cycle
\cite{Frachal}:

\begin{eqnarray}
x''&=&x'+x \nn \\
p''&=&p'+p \\
\zeta''&=&\zeta'\zeta e^{-\frac{i}{\hbar}xp'} \nn
\end{eqnarray}

\ni from which we derive the left- and right-invariant vector fields:

\be
\ba{lcl}
\XL{x} & = & \parcial{x} - \frac{1}{\hbar}p\,\Xi  \\
\XL{p}& = & \parcial{p}  \\
\XL{\z} & = & i\z\parcial{\z} \equiv \Xi  
\ea \,\,\,\,\,\,
\ba{lcl}
\XR{x}& = & \parcial{x}  \\
\XR{p}& = & \parcial{p} -  \frac{1}{\hbar}x\,\Xi  \\
\XR{\z} & = & i\z\parcial{\z} \equiv \Xi 
\ea \label{XLRH-Wc}
\ee

\ni and a polarization subalgebra generated by ${\tilde{X}^L}_p$ can
be chosen.

Starting from a complex function $\Psi(x,p,\zeta)$ on the group, the
$U(1)$-function condition just says that $\Psi(x,p,\zeta)=\zeta\psi(x,p)$,
whereas the $\check{T}$-function condition establishes the quasi-periodicity
condition:

\begin{equation}
\psi^{\epsilon}(x+kL,p)=e^{\frac{i}{\hbar}\epsilon kL}\psi^{\epsilon}(x,p)\;\; .
\end{equation}

\ni Applying the polarization condition we get:

\begin{eqnarray}
\psi^{\epsilon}(x,p)&=&\Phi^{\epsilon}(x) \\
\Phi^{\epsilon}(x+L)&=&e^{\frac{i}{\hbar}\epsilon L}\Phi^{\epsilon}(x)\;\; , \nn
\end{eqnarray}

\ni so that, the action of the {\it quantum operators} is

\bea
\hat{p} \Psi^\ep & = & \z \nabla\Phi^\ep \nn   \\
\hat{x} \Psi^\ep & = & \z x \Psi^\ep \;\; . \label{H-WcOConf}
\eea

We can see that only $\hat{p}$ and {\it finite} boosts by an amount of
$p_k\equiv\frac{2\pi\hbar}{L}k$ are {\it good} operators. Boosts in general
{\it change the quantization index} $\epsilon$. Then, the question arises
of up to whether or not a good position-like operator can be found.
To see this question, we realize that the classical function $\eta\equiv
e^{i\frac{2\pi}{L}x}$ is periodic and, therefore, $\hat{\eta}^k\equiv
(e^{i\frac{2\pi}{L}\hat{x}})^k$ is a {\it good} position operator for $k\in Z$,
although $\hat{\eta}$ is not Hermitian in general (it is rather unitary).
However, the operator $\hat{\eta}$ can be decomposed in two Hermitian
pieces (the sum of one Hermitian and other anti-Hermitian more precisely):

\begin{equation}
\hat{\eta}=\cos(\frac{2\pi}{L}\hat{x})+i\sin(\frac{2\pi}{L}\hat{x})\;\; .
\end{equation}

Finally, it is worth mentioning that $<\hat{p},\hat{\eta},\hat{\eta}^{\dag}>$
is a set of good operators which closes, under ordinary commutation, a
non-extended oscillator algebra. The operators $\hat{\eta}$ and ${\hat{\eta}}^{\dag}$
act as ladder operators on the eigenfunctions of $\hat{p}$, a fact which 
has been used in \cite{O-K} to study Quantum Mechanics on the circumference.

\end{document}